\documentclass[manuscript]{aastex}
\begin{document}

\title{A statistical study of the post-impulsive-phase acceleration of flare-associated coronal mass ejections}

\author{X. Cheng\altaffilmark{1}, J. Zhang\altaffilmark{2,1}, M. D. Ding\altaffilmark{1}, and W. Poomvises\altaffilmark{2}}

\altaffiltext{1}{Department of Astronomy, Nanjing University,
Nanjing, 210093, China; dmd@nju.edu.cn} \altaffiltext{2}{Department
of Computational and Data Sciences, George Mason University, 4400
University Drive, MSN 6A2, Fairfax, VA 22030}

\begin{abstract}
It is now generally accepted that the impulsive acceleration of a
coronal mass ejection (CME) in the inner corona is closely
correlated in time with the main energy release of the associated
solar flare. In this paper, we examine in detail the
post-impulsive-phase acceleration of a CME in the outer corona,
which is the phase of evolution immediately following the main
impulsive acceleration of the CME; this phase is believed to
correspond to the decay phase of the associated flare. This
observational study is based on a statistical sample of 247 CMEs
that are associated with M- and X-class GOES soft X-ray flares from
1996 to 2006. We find that, from many examples of events, the CMEs
associated with flares with long-decay time (or so-called
long-duration flares) tend to have positive post-impulsive-phase
acceleration, even though some of them have already obtained a high
speed at the end of the impulsive acceleration but do not show a
deceleration expected from the aerodynamic dragging of the
background solar wind. On the other hand, the CMEs associated with
flares of short-decay time tend to have significant deceleration. In
the scattering plot of all events, there is a weak correlation
between CME post-impulsive-phase acceleration and flare decay time.
The CMEs deviated from the general trend are mostly slow or weak
ones associated with flares of short-decay time; the deviation is
caused by the relatively stronger solar wind dragging force for
these events. The implications of our results on CME dynamics and
CME-flare relations are discussed.

\end{abstract}


\section{Introduction}

Coronal mass ejections (CMEs) are large-scale solar activities,
which can release a vast amount of plasma and magnetic flux into the
outer space and cause interplanetary disturbances and geomagnetic
storms near the Earth \citep{gos93,webb94}. Flares are viewed as
strong energy release in the lower atmosphere of the Sun, where CMEs
originate from but then depart from the Sun. The physical
relationship between CMEs and flares has been a long-standing
elusive issue in solar physics \citep{kahler92,gos93,hund99}.
Nevertheless, recent studies demonstrate that there is a strong
physical connection between CMEs and flares. Zhang et al. (2001,
2004) studied the whole kinematic process of CMEs and found that
those CMEs associated with flares usually undergo three distinct
phases of evolution: the initiation phase, impulsive acceleration
phase (mainly in the inner corona, $\leq$ 3.0 $R_\odot$), and
propagation phase (mostly in the outer corona). Furthermore, it was
found that the three kinematic phases of CMEs coincide in time very
well with the three phases of the associated flares: the pre-flare
phase, flare main energy release phase or rise phase in soft X-ray,
and flare decay phase, respectively \citep{zhang01,bur04,vr05b}.
Recently, Temmer et al. (2008) analyzed the kinematics of two fast
halo CMEs in the inner corona and found that there was a close
connection between the acceleration profiles of the CMEs and the HXR
light curves of the related flares. The almost synchronized temporal
correlation between CME acceleration and flare flux increase
indicates that both are driven by the same energy release process in
the corona, especially within the impulsive phase. In other words,
the dynamic evolution of these two phenomena may be different
manifestation of the same energetic process, presumably via magnetic
reconnection \citep{lin00,priest02,vr04b,zhang06,mari07,temmer08}.
Therefore, there is no apparent cause-effect relation between them,
i.e., they do not cause one another.

In statistical views, the more intensive the flares are, the greater
the possibility of the flares being associated with CMEs is
\citep{and03,ya05}. Yashiro et al. (2006) studied the power-law
indices of the frequency distribution of flares, and found that
flares with CMEs have a harder index of distribution than that of
flares without CMEs. Zhang et al. (2003) found that flares
associated with fast CMEs show clear footpoint-separating and
two-ribbon brightening, while this feature is less often in flares
associated with slow CMEs or without CMEs. MacQueen and Fisher
(1983) and recently by St. Cyr et al. (1999) found that CMEs
associated with flares or active regions have relatively higher
speeds and tend to propagate with a constant speed or a negative
acceleration in the outer corona; while ones associated with
eruptive filaments have an initial slow speed and a positive
acceleration in the outer corona. Using the latest CDAW CME
catalog\footnote{http://cdaw.gsfc.nasa.gov/CME\_list}, Moon et al.
(2002) also found similar results.

Nevertheless, the detailed relationship between the CME evolution
following the impulsive acceleration phase and the properties of the
flare decay phase has not been studied. Prior to LASCO, CMEs were
commonly thought to propagate with nearly constant speed. Now we
know that CMEs usually have a small acceleration or deceleration in
the outer corona (about between 3.0 and 30.0 $R_\odot$) after they
have been strongly accelerated in the inner corona
\citep{and01,neu01,zhang01,gall03,sha03}. But, how this late
evolution, dubbed as ``post-impulsive-phase acceleration", is
related with flare characteristics is not clear, and therefore a
detailed study on this issue is desirable. To quantify such
evolution of CMEs, we introduce a fixed time window of two hours
beginning at the peak time of the associated flare to calculate the
acceleration (details of methods given in the next section). It is
noted that the post-impulsive-phase acceleration is likely related
to the residual acceleration originally proposed by Chen et al.
(2003). Based on their theoretical flux-rope model, Chen et al.
(2003) specified the residual acceleration (in differ from the main
acceleration) to the acceleration of the period that Lorentz
self-force is decreased and the dragging force of solar wind starts
to dominate. In the observational context, the residual acceleration
was used by Zhang et al. (2006) in a more general sense to refer to
the observed velocity change of CMEs following the impulsive
acceleration phase, which can be practically separated by the peak
time of the associated soft X-ray flares. In this paper, we
investigate the post-impulsive-phase acceleration through both a
case study of a variety of typical events and a statistical study as
well. The main finding is that CMEs associated with long-decay
flares tend to have positive post-impulsive-phase acceleration, and
thus are more likely to reach a higher peak speed. In $\S2$, we
present the observations. Example events of diversified properties
are presented in $\S3$. Our statistical results on
post-impulsive-phase accelerations are shown in $\S4$. The more
general relations between CMEs and flares are given in $\S5$
followed by a summary and discussions in $\S6$.
\section{Observations and Data}

In this study, we make use of CMEs observed during 1996--2006 by the
the Large Angle and Spectrometric Coronagraph (LASCO) \citep{bru95}
on board the Solar and Heliospheric Observatory (SOHO). The two
complementing LASCO coronagraphs, C2 and C3, have fields of view
(FOVs) of 2.2--6.0 $R_\odot$ and 4.0--30 $R_\odot$, respectively.
Flare data are from  GOES satellites providing the full disk soft
X-ray emission from the Sun in 1--8 {\AA}. The RHESSI (Reuven Ramaty
High Energy Solar Spectroscopic Imager; Lin et al. 2002) and YOHKOH
SXT (Soft X-ray Telescope) provide the hard X-ray light curves for
some of the flares studied in this paper. However, the hard-X-ray
flare data do not enter the statistical study in this paper.

From 1996--2006, there are in total about 11536 CMEs observed by
LASCO according to the CDAW CME catalog, and 22686 flares seen by
GOES based on the NOAA flare catalog. Because of the sheer number,
we limit our study only to major flares; these are 1425 M-class and
120 X-class flares, and 1545 in total. In order to find out only
those flares associated with CMEs (the so-called eruptive flares),
an easy and quick approach is to use the so called time-window
method, without resorting to inspecting images (Harrison 1995,
Yashiro et al. 2005). The CME onset time is estimated through a
backward extrapolation to the surface at 1.0 $R_{\sun}$ from the
height-time observations in coronagraph assuming a constant
velocity. An association is assumed if a flare occurs within a
certain time window centered at the estimated CME onset time, e.g.,
$\pm$ 60 minutes. We find that this simple time-window method can
make successful association for most of events ($\sim$85\%).
However, there is a certain percentage of wrong association, which
may not be acceptable for serious studies, e.g., the work presented
in this paper and predicting CMEs from flare observations. The wrong
association arises from the chance association, e.g., between a
confined solar flare occurring on the front-side of the Sun and a
CME occurring on the back-side of the Sun.

In this paper, we use a more strict method to associate flares and
CMEs, by visually inspecting the movies observed by LASCO with the
those by EIT (Extreme-ultraviolet Imaging Telescope, also on board
SOHO, Delaboudini\`{e}re et al. 1995) one by one, although it is
tedious and time consuming. In addition to be used to identify the
location of flares through transient brightening in a small compact
patch, EIT data are also commonly used to identify the source region
of CMEs through the signature of large scale dimming and/or wave,
which are often prominent for major eruptions. The CME is taken for
being associated with flare if the temporal-spatial co-registration
of transient flare brightening and the large scale dimming on EIT
images occurs, which is the most reliable way to associate flares
and CMEs.

Among the 1545 major solar flares recorded by NOAA, 1246 events have
both EIT and LASCO observations; the other 299 flares occurred in a
period of either EIT or LASCO data gap (or both). For these 1246
flares, we find that 706 events (56.6\%) are associated with CMEs,
while the other 540 flares (43.4\%) are confined. These confined
flares will not be used in this study. Further, we eliminate those
events with less than 5 effective snapshot observations during the
two-hour window of calculating the post-impulsive-phase acceleration
(explained in the next paragraph). We also remove those events
without effective C2 observations. In the end, we obtain 247
flare-CME pairs suitable for the study in this paper.

To determine the magnitude of the post-impulsive-phase acceleration,
we calculate the average acceleration within the fixed time window
of two hours beginning at the peak time of the associated flare.
While the post-impulsive-phase refers to the CME evolution, we have
adopted the flare peak time as the proxy of the starting time of
this CME phase. The starting time is difficult to be determined
directly from CME observations, due to the poor cadence and the lack
of inner coronal observations of LASCO (except for a small number of
events). Further, we believe that this proxy is a reasonable one
because of the temporal coincidence between CME kinematic evolutions
and flare flux variations (e.g., Zhang et al. 2001). The average
acceleration is obtained through the second-order polynomial fitting
of the observed height-time measurements in this window. We believe
that such an average acceleration value is an effective
representation of the post-impulsive-phase acceleration of the CME,
whereas its validity may need to be further demonstrated using
observations with higher cadence. As for the method of the fixed
time window, we think that it is a good approach to characterize the
relevant observation. In this approach, the beginning time of the
post-impulsive-phase acceleration phase is uniform and well defined,
that is the peak time of the associated flare. On the other hand,
for this type of observational study of calculating an average
property from a limited number of data points, one has to choose the
most appropriate window. We think that the selection of a two-hour
window is reasonable; it is long enough to have sufficient number of
data points to make a second-order polynomial fitting to the CME
height-time measurement, while is short enough to differentiate the
coronal effect on the dynamic evolution from the otherwise dominant
solar wind effect on CME evolution in the later phase of the
propagation in the LASCO FOV. Furthermore, we have checked the
influence of the different time window on the data and find that
there is always a weak correlation between the CME
post-impulsive-phase acceleration and the associated-flare decay
time. Therefore, similar results are obtained if a different time
window is adopted.


Note that the average acceleration through the fixed time window
carries much smaller error than the uncertainty inferred from any
piece-wise fitting method. In general, the acceleration is much more
difficult to calculate than the speed due to the nature of
differentiation on discrete data points; the error bars in the
acceleration values are significantly larger than those in the speed
values \citep{zhang06,ya04}. For one specific data point, the
acceleration error could be comparable to the inferred acceleration
value. However, the error in the average acceleration based on the
second-order polynomial fitting of at least five observed
height-time measurements becomes significantly smaller.

\section{Examples} \label{bozomath}
Before we show the statistical properties of CME
post-impulsive-phase acceleration and flare decay time, we present
in detail four individual events, two of which are of positive
acceleration and the other two are of negative acceleration. The
overall properties of these four events are summarized in Table 1,
including CME velocity, impulsive acceleration, post-impulsive-phase
acceleration, acceleration error, flare rise time, decay time,
location and peak intensity.
\subsection{CME Positive Post-Impulsive-Phase-Acceleration and Flare Long-Decay Time}
\subsubsection{2001 September 24 Event}
The CME on 2001 September 24 is associated with a GOES X2.6 class
flare. Figure 1 shows the velocity-time plot of the CME (broken
lines with symbols) along with the GOES X-ray time profiles (solid
line). The CME onset time, estimated through linear extrapolation of
the height-time measurement, was at 10:21 UT while the associated
flare started at 09:32 UT; there is an 49 minute difference between
the two onset times. We argue that, if inner corona observation were
available, one would expect to find that the CME onset time coincide
with the flare onset, probably within a few minutes. The time
difference we find here is due to the usage of linear extrapolation
assuming a constant speed in the inner corona, which is apparently
an over simplification of the true evolution involving significant
acceleration from almost zero speed to the final speed (e.g., Zhang
et al. 2001). The heliographic location of the flare was S16\,E23,
which is consistent with the CME feature position angle of
142$^{\circ}$. The feature position angle is defined as the most
distinguishable feature used to measure the height. Here, we use the
feature position angle instead of the center position angle because
the CME is a halo CME when it appears in the FOV of LASCO/C3. The
CME first appeared in the FOV of the C2 image at 10:30 UT at a
height of 3.3 $R_\odot$ from the disk center. As shown in Figure 1,
the CME reached a velocity at about 2000 km s$^{-1}$ at the peak
time of the flare. What is important of this event is that it
continued to accelerate during two hours after the peak time of the
flare, from about 2000 to 2500 km s$^{-1}$; during this period, the
CME leading edge moved from about 3 $R_\odot$ to 25 $R_\odot$. The
post-impulsive-phase acceleration during the fixed time window is
about 64 m s$^{-2}$, as determined by the second-order polynomial
fitting of the height-time measurements. On the other hand, the
inferred CME impulsive acceleration during the impulsive
acceleration phase is about 455 m s$^{-2}$, using the flare rise
time as a proxy of the time of CME impulsive acceleration (Zhang et
al. 2006). Note that the uncertainty of the CME speed comes mainly
from the uncertainty in height measurements, which are estimated to
be about 8 pixels in the original images, or about 0.10 and 0.47
$R_\odot$ for C2 and C3, respectively. In the same way, the
uncertainty in height measurements also determined the acceleration
uncertainty, which is $\pm$58 m s$^{-2}$ for this event. The same
uncertainties are used for other events discussed in this paper. We
note that the uncertainties may be even larger than such determined
for CME events with less sharp leading edges. The flare peaked at
10:38 UT and ended at 11:09 UT with a decay time of 31 min; the
ending time is defined by NOAA as the time of the half-maximum. The
flare is apparently a long decay flare as seen from the temporal
profile of the SXR emission in Figure 1. The decay phase lasts
longer than the radiation cooling time scale (about 20 minutes), so
there must be the continuing energy released to delay the radiation
cooling time. We believe that the observed post-impulsive-phase
acceleration of this CME is related with the continuing energy
release following the impulsive energy release phase known for such
a long-decay flare. Continuing driving force is needed, not only to
overcome the aerodynamic dragging force of the background solar
wind, but also to further accelerate the CME.

\subsubsection{2003 November 18 Event}
The CME on 2003 November 18 is another example of events showing
positive post-impulsive-phase acceleration. It is a CME associated
with a GOES M4.5 class flare (Figure 2). The estimated CME onset
time from linear extrapolation is 9:43 UT and the associated flare
started at 9:23 UT. The heliographic coordinate of the flare was
S14\,E89 and the feature position angle of the CME that was about
87$^{\circ}$. The CME first appeared in the FOV of C2 at 9:50 UT at
a height of 2.9 $R_\odot$. As from Figure 2, the CME was
continuously accelerated from about 1300 to 1900 km s$^{-1}$ during
two hours after the peak time of the flare. The post-impulsive-phase
acceleration and its error are about 40 m s$^{-2}$ and $\pm$36 m
s$^{-2}$, respectively. On the other hand, the estimated impulsive
acceleration during the impulsive acceleration phase is about 632 m
s$^{-2}$. The flare peaked at 10:11 UT and ended at 11:01 UT with a
decay time of 50 min, which implies a long decay behavior of the
flare.
\subsection{CME Negative Post-Impulsive-Phase-Acceleration  and Flare Short-Decay Time}

\subsubsection{2004 October 20 Event}
Different from the previous two events, we show examples of CMEs
with negative post-impulsive-phase acceleration, or deceleration
following the impulsive acceleration phase. In Figure 3, we shown a
CME that occurred on 2004 October 20 and was associated with a GOES
M2.6 class flare. The CME onset time was 10:32 UT and the associated
flare started at 10:43 UT. The position of the flare was N11\,E68.
The center position angle and width of the CME were 68$^{\circ}$ and
123$^{\circ}$, respectively. It is evident that the CME was
decelerated from about 1100 to 700 km s$^{-1}$ during two hours
after the peak time of the flare, as seen from the velocity profile
of the event in Figure 3. The post-impulsive-phase acceleration and
its error of the CME are about --71 m s$^{-2}$ and $\pm$37 m
s$^{-2}$ in the fixed time window, respectively. The flare peaked at
10:51 UT and ended at 10:56 UT with a decay time of only 5 min. This
event indicates that a CME, when is associated with a short decay
flare, lacks the continuing driving force and therefore suffer
significant deceleration after the impulsive acceleration phase.

\subsubsection{2005 August 25 Event}
The following is another example of CMEs of negative
post-impulsive-phase acceleration. The fast CME was observed on 2005
August 25 and was associated with a GOES M6.4 class flare. The CME
onset time was 4:16 UT and the associated flare started at 4:31 UT.
N09\,E80 was the site of the flare. The center position angle and
width of the CME were 75$^{\circ}$ and 146$^{\circ}$, respectively.
The LE of the CME was very sharp and can be easily identified from
the running-difference images. It had a velocity of about 2000 km
s$^{-1}$ when appeared in the FOV of C2. The velocity profile of
this CME in Figure 4 shows that the CME was decelerated from about
2000 to 1400 km s$^{-1}$ in about 2 hours. The post-impulsive-phase
acceleration and its error of the CME are --129 m s$^{-2}$ and
$\pm$41 m s$^{-2}$, respectively. The flare peaked at 04:40 UT and
ended at 04:45 UT. The decay phase of the flare lasted only 5 min in
the SXR temporal profile.

\section{Statistics on CME Post-impulsive-phase Acceleration and Flare Decay Time} \label{bozomath}

Having presented examples of two distinct types of events, we now
look into the statistical behavior. Figure 5 shows the distribution
of the post-impulsive-phase acceleration of CMEs vs. the decay time
of the associated flares for all the 247 events studied. As shown by
the linear fitting line in Figure 5, there is a general trend that
the longer the decay time of flares, the larger the
post-impulsive-phase acceleration of CMEs. On the other hand, the
correlation between CME post-impulsive-phase acceleration and flare
decay time is rather poor. There is an apparent-wide scattering of
the parameters in both dimensions. The acceleration varies in a wide
range from --150 m s$^{-2}$ to 180 m s$^{-2}$ and tends to change
from negative to positive value with the decay time of the
associated flares increasing. The fraction of CME events with
positive post-impulsive-phase acceleration apparently increases as
the decay time of the associated flare increases. The mean
post-impulsive-phase acceleration of all events is negative at about
--11.9 m s$^{-2}$, as indicated by the solid thin line in the
Figure.

For the sake of clarity of discussion, we divide the events into
three different groups, as enclosed by the three rectangular boxes
in the Figure: positive-post-impulsive-acceleration CMEs with
long-decay flares (acceleration-long, or A-L),
negative-post-impulsive-acceleration CMEs with short-decay flares
(deceleration-short, or D-S), and
positive-post-impulsive-acceleration CMEs with short-decay flares
(acceleration-short, or A-S). Apparently, there is a lack of CME
events of deceleration with long-decay flares; those CMEs associated
with long decay flares tend to have positive post-impulsive-phase
acceleration, even though they have reached a high speed at the end
of the impulsive acceleration. It is commonly believed that there is
a continuing energy release during the decay phase of long duration
flares. Therefore, this statistical result implies that the
continuing energy release also related with the continuing
acceleration of CMEs in the outer corona.

Nevertheless, for events with short flare decay times, the
post-impulsive-phase acceleration could be either positive or
negative. One would expect that the post-impulsive-phase
acceleration tends to be negative for events associated with
short-decay flares, since there is no further energy available for
accelerating CMEs following the main energy release phase. However,
we believe that the influence of solar wind dragging force makes the
matter complex. It is known that the solar wind dragging force
accelerates slow CMEs and decelerate fast CMEs; the magnitude
depends on the relative velocity between CME and solar wind, the
drag coefficient and the CME cross section size \citep{cargill04}.
Indeed, we identify that many positive-acceleration-short-decay
events (A-S) are slow CMEs. These slow CMEs also tend to be narrow
and weak. Because of the slowness of these events, the positive
acceleration is likely to be caused by the dragging force of the
background solar wind, instead of the continuing energy release that
only occurs in A-L type events. Further, slow and narrow CME events
often appear weak in brightness and thus fuzzy in morphology as seen
in coronagraph images. As a consequence, it is usually hard to trace
a consistent features such as leading edges (LE) for these event,
leading to large error in the height-time measurements and thus
larger error in the derived velocity.

In Figure 6, we show a similar scattering plot as in Figure 5 but
use only events with fast CME speed ($>$800 km s$^{-1}$) and wide
CME angular width ($>$60$^{\circ}$). The general trend between CME
post-impulsive-phase acceleration and flare decay time becomes more
distinct than that in Figure 5. In essence, we want to argue that,
for major fast and wide CMEs, the CMEs associated with long
decay-time flares tend to be further accelerated in the outer
corona, overcoming the slowing-down effect of solar wind dragging on
fast CMEs. The CMEs associated with short decay-time flares tend to
be decelerated in the outer corona, but may gain positive
acceleration due to the solar wind dragging if the initial speed is
slow.

\section{General Statistical Relations Between CME and Flare Properties} \label{bozomath}
In this section, we describe the more general relations between CMEs
and flares from a statistical point of view. In Figure 7, we show
the scattering plots between CME velocity and various flare
parameters, including rise time, total duration, peak flux and total
flux (or fluence). We find that the velocities of CMEs show almost
no correlation or weak at best with the rise times or the total
durations of the associated flares. However, there is a certain
positive correlation between CME velocity and flare peak flux. This
correlation has been noted before \citep{moon02,moon03}. The
linear-fitting formula between CME velocity $V$ (km s$^{-1}$) and
flare peak flux $F$ (Watt m$^{2}$ s$^{-1}$) can be expressed as $V =
474.0 log F + 2922.2$.

It is interesting that the best correlation is found between CME
velocity and total flare flux or fluence; the total flux here is
simply calculated through the product of the peak flux and the total
duration of flares. The linear correlation coefficient is 0.54,
which is better than that of any other flare parameters. The
linear-fitting formula between CME velocity and flare fluence can be
expressed as $V = 422.4 log (F \cdot T) + 1321.6$, where $F$ is the
peak flux and $T$ is the total duration in second. The total soft
X-ray flux of a flare is believed to be a good measure of the total
energy released during the flare. Therefore, this correlation result
may be a manifestation that the final CME velocity is proportional
to the total energy released in the corona. During the eruption
process, part of the released coronal energy goes to CME bulk
kinetic energy in the global scale, while the other part goes to
flare plasma heating and particle acceleration in the microscopic
scale. Magnetic reconnection is a likely physical process that
produces both CME and flare energies, and also makes the amounts of
the two energies comparable.

Figure 8 shows the relationship between the velocity and the angular
width of CMEs. Apparently, there is a positive correlation; the
correlation coefficient is 0.52 for all 247 CMEs and a much improved
coefficient of 0.62 for 111 limb CMEs (heliographic longitude larger
than 60$^{\circ}$). It is not surprising that the limb CMEs show
better correlation, since they are less subject to the project
effect that artificially enlarges the apparent angular width. The
linear-fitting formula for the limb CMEs is $V = 2.8 W + 429.1$,
where $W$ is the apparent CME angular width in units of degree. The
correlation coefficient obtained here is larger than that obtained
for a larger sample of events by Yashiro et al. (2004)

Finally, we show the histogram distribution of the CME/flare
parameters used in this study (Figure 9). Their overall statistical
properties, including minimum value, maximum value, average value,
medium value, mode (or the value at maximum distribution) and
standard deviation are summarized in Table 2. The mean velocity of
747 km s$^{-1}$, for the 247 major events studied in this paper, is
larger than the value of 684 km s$^{-1}$ obtained by Moon et al.
(2002). The mean impulsive acceleration of these CMEs is 939 m
s$^{-2}$, whereas the post-impulsive-phase accelerations are limited
to a small range centered near zero and the mean
post-impulsive-phase acceleration is --11.9 m s$^{-2}$, being almost
consistent with the results of Zhang et al. (2006).

\section{Summary and Discussions}
We have studied the statistical kinematic properties of
major-flare-associated CME events occurred between 1996-2006, with a
focus on the post-impulsive-phase acceleration. It has been well
known that a typical flare-associated CME usually has a strong and
impulsive acceleration in the inner corona, and the impulsive
acceleration coincides well with the main energy release of the
associated flare \citep{zhang01,gall03,vr04b,temmer08}. In this
paper, we further find that the post-impulsive-phase acceleration of
CMEs may be also physically related with the continuing energy
release. In particular, CMEs tend to have positive
post-impulsive-phase acceleration if the associated flares have a
long decay phase; this kind of flares is often called LDE (long
duration event). When flare decay times are short, accompanying CMEs
may have mixed response, possibly with positive or negative
post-impulsive-phase acceleration from event to event.

We argue that the positive post-impulsive-phase acceleration of LDE
events is driven by the continuing magnetic reconnection occurring
during the flare decay phase. It is widely accepted that the main
energy release phase of a solar flare is driven by magnetic
reconnection in the inner corona. When a flare has a long decay
phase, continuing magnetic reconnection is also needed to explain
the lasting thermal emission. Without the continuing energy release,
the typical thermal energy decay time, which is controlled by the
radiation cooling and also the thermal conduction to the cooler
chromosphere, is only about 20 minutes \citep{forbes89,isobe02}. The
idea of continuing reconnection has also been supported by
observations of long-lasting post-flare loops
\citep{schmieder95,schmieder96,czay99,shee04,kolo07}. The observed
rising motion of post-flare loops, as well as the observed
separation motion of flare ribbons, are well explained by the
reconnection model that these observed motions are driven by
systematic rising of the reconnection point in the corona. The
observed down-flows above post-flare arcades in the long-duration
flares also support the idea of continuing reconnection
\citep{mckenzie00,shee04}. Similar connection between ribbon
separation motion and coronal magnetic reconnection also occur in
the main energy release phase \citep{qiu04,qiu05}. Nevertheless,
there must be certain differences between the reconnection during
the main energy release phase and that during the decay phase. Isobe
et al. (2002) found that in the decay phase the reconnection rate,
and also the energy release rate, was about one-tenth of that in the
rise phase. Therefore, it is reasonable to argue that the impulsive
CME acceleration is driven by the fast reconnection in the impulsive
phase, while the post-impulsive-phase CME acceleration is caused by
the slow reconnection after the impulsive phase.

Chen et al. (2000, 2003, 2006) devised an analytic flux rope model
to explain CME's main and residual accelerations. It seems that the
residual acceleration in the context of Chen's model is similar to
the post-impulsive-phase acceleration discussed in this paper,
especially in terms of their magnitude and timing relative to that
of the main phase. They showed that the main acceleration is
attained before the CME reaches a critical height (below 2--3
$R_\odot$), and is then followed by the residual acceleration. The
main acceleration phase is dominated by the Lorentz self-force
through the injection of the poloidal magnetic flux of the flux
rope, while the residual acceleration is dominated by the solar wind
aerodynamic dragging force \citep{chen03}. On the other hand, in the
scenario of magnetic reconnection models, the CME impulsive
acceleration is driven by the fast reconnection. It is kind of
runaway tether-cutting reconnection not only cutting off the field
lines tied to the photosphere and lessening the restraint of the
overlying field but also rapidly increasing the magnetic pressure
below the flux rope due to an added poloidal flux
\citep{moore01,zhang06}.
When the runaway tether-cutting reconnection ceases, the impulsive
acceleration stops. However, for the CME associated with a long
decay flare, even though it has departed from the Sun and may have
already moved to as far as several solar radii from the Sun, the
reconnection process may continue; it is most evident in the
formation of the post-eruption loop arcades. The continued
reconnection may further drives the CME and thus produce the
positive post-impulsive-phase acceleration of the CME.
Note that, the difference between the impulsive acceleration and the
post-impulsive-phase acceleration may be mainly their different
reconnection rates. A question may arise with respect to how to
connect the positive post-impulsive-phase acceleration in the outer
corona with the reconnection site close to the surface of the Sun. A
possible explanation is that the reconnection magnetic fields are
overlying fields surrounding the CME leading edge in the outer
corona but are stretched up open in the low corona. Therefore, the
post-impulsive-phase acceleration of the CME in the outer corona and
the energy release of the associated flare in the decay phase are
different manifestations of the slow magnetic reconnection following
the fast magnetic reconnection.

On the other hand, when the associated flare is of short duration,
one does not expect continuing reconnection and thus continuing
acceleration of the CME. Indeed, about half CMEs with short-duration
flares have suffered deceleration in the outer corona, or negative
post-impulsive-phase acceleration. Nevertheless, many CMEs
associated with short-duration flares have also showed positive
post-impulsive-phase acceleration. Detailed investigation shows that
these CMEs tend to be slow, narrow and weak. It is likely that the
positive acceleration is caused by the solar wind dragging force,
which acts as a positive driving force when the embedded object is
slower. The full dynamic evolution of a CME shall involve not only
the various Lorentz forces caused by current-carrying magnetic
fields, but also the solar wind dragging force. For those fast CME
events associated with long decay flares, even though the solar wind
dragging force is a decelerator of CMEs, the positive Lorentz force
caused by continuing magnetic reconnection is able to overcome the
dragging and further accelerating the already-fast CMEs. This
scenario helps explain why certain number of CMEs are extremely
fast.

\acknowledgments

We thank the anonymous referee for many valuable suggestions and
comments that helped to improve the paper significantly. We are
grateful to P. F. Chen and M. Jin for discussions and help in data
analysis. J. Z. thanks Seji Yashiro for providing the text file of
the CDAW CME catalog and his version of CME-flare list. J. Z. and W.
P. acknowledge the support from NSF grant ATM-0748003 and NASA grant
NNG05GG19G. X. C. and M. D. D. are supported by NSFC under grants
10673004, 10828306, and 10933003 and by NKBRSF under grant
2006CB806302. We acknowledge the use of the CDAW CME catalog, which
is generated and maintained at the CDAW Data Center by NASA and The
Catholic University of America in cooperation with the Naval
Research Laboratory. SOHO is a project of international cooperation
between ESA and NASA. Yohkoh is a project of international
cooperation between Janpan, US and UK.

\clearpage

\begin{figure}
\epsscale{1.00} \plotone{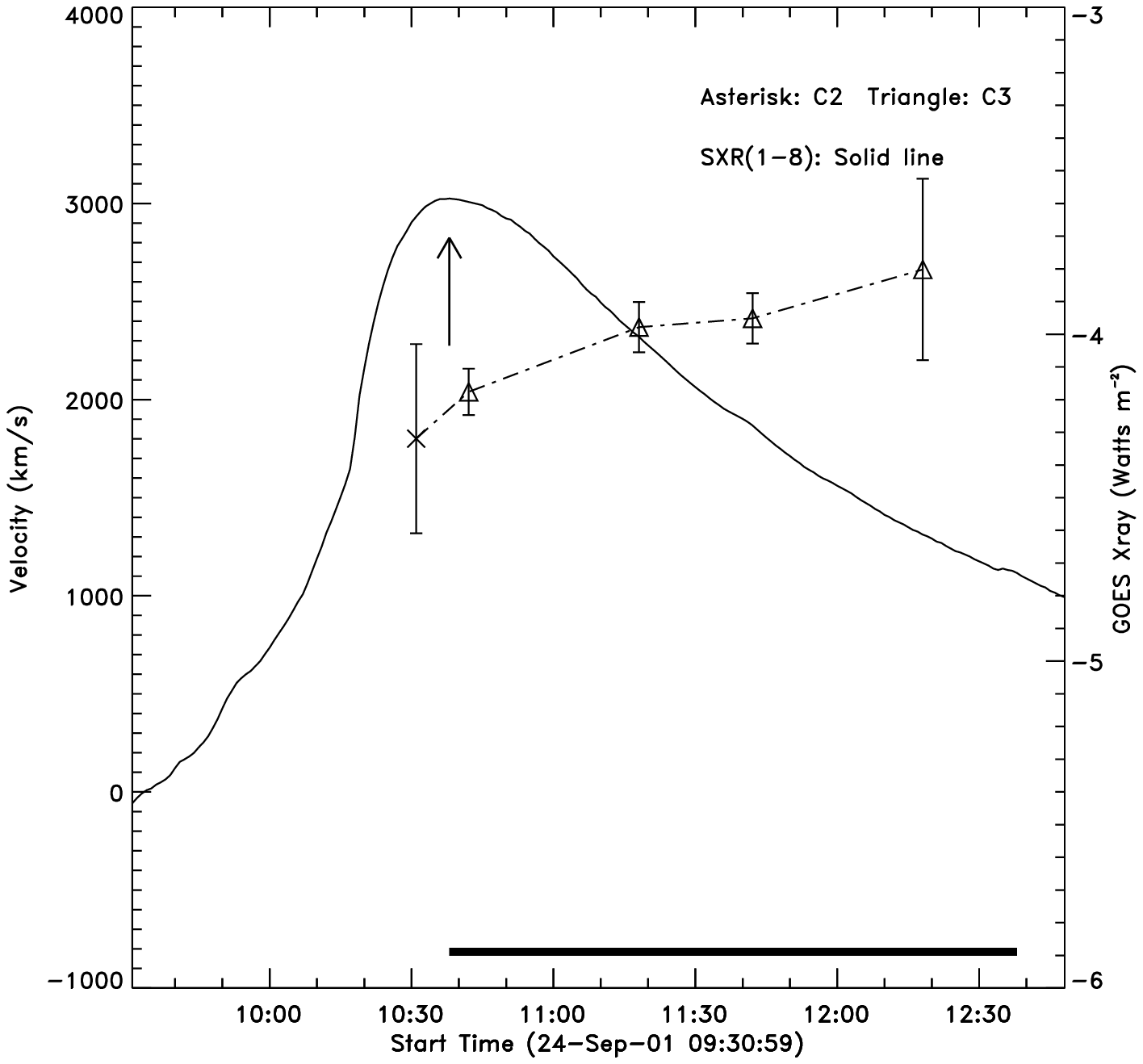} \caption{The velocity
evolution of 2001 September 24 CME (broken line with symbols) and
the temporal profile of the GOES SXR flux of the associated flare
(solid line); the arrow denotes the peak time of the flare and the
onset of the post-impulsive-phase acceleration of the CME. The
horizontal bar indicates the two-hour window used to calculate the
post-impulsive-phase acceleration. The CME has a positive
post-impulsive-phase acceleration and the flare is of
long-decaying.\label{fig1}}
\end{figure}

\begin{figure}
\epsscale{1.00} \plotone{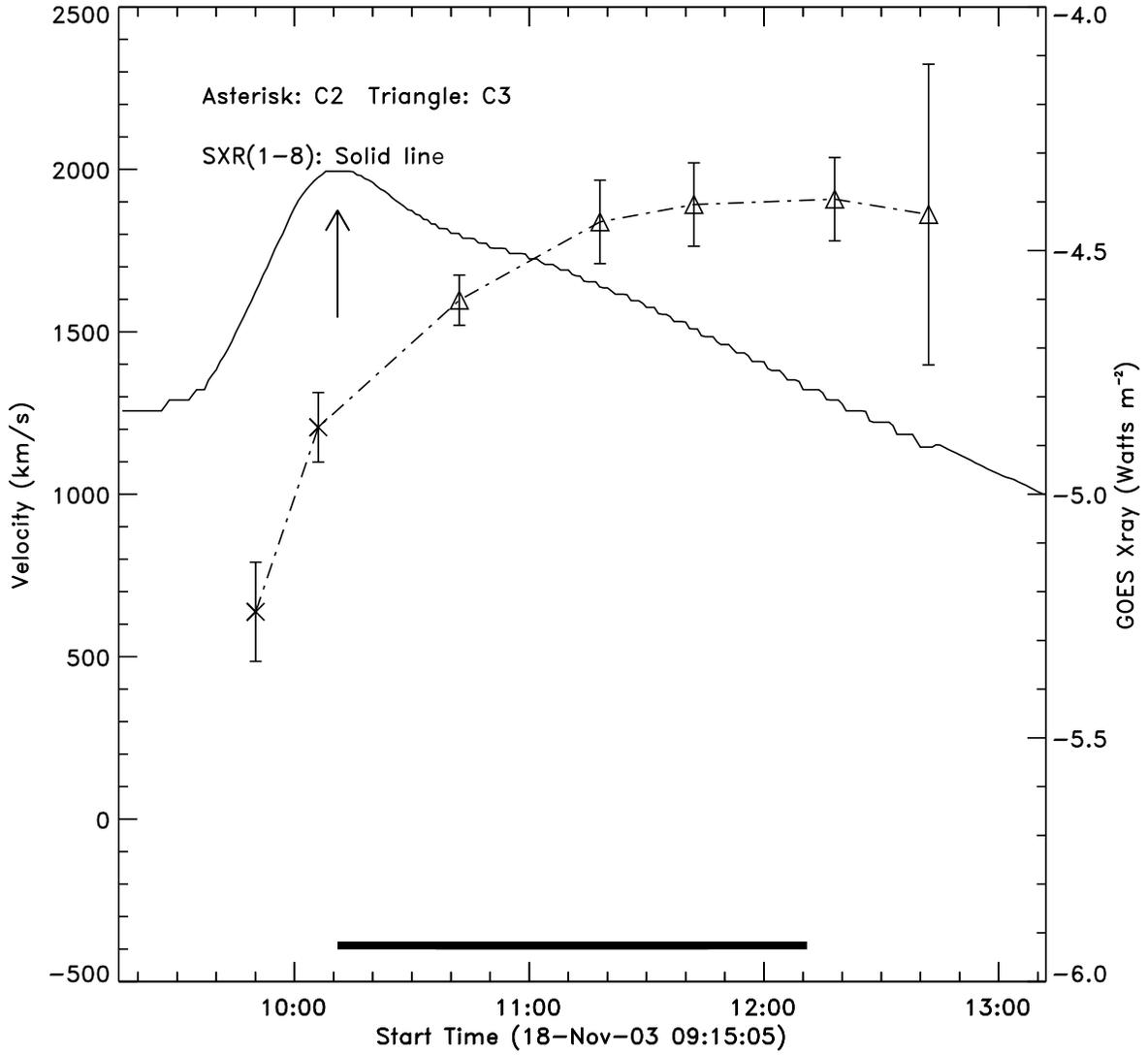} \caption{The velocity
evolution of 2003 November 18 CME (broken line with symbols) and the
temporal profile of the GOES SXR flux of the associated flare (solid
line). The CME has a positive post-impulsive-phase acceleration and
the flare is of long-decaying.\label{fig2}}
\end{figure}

\begin{figure}
\epsscale{1.00} \plotone{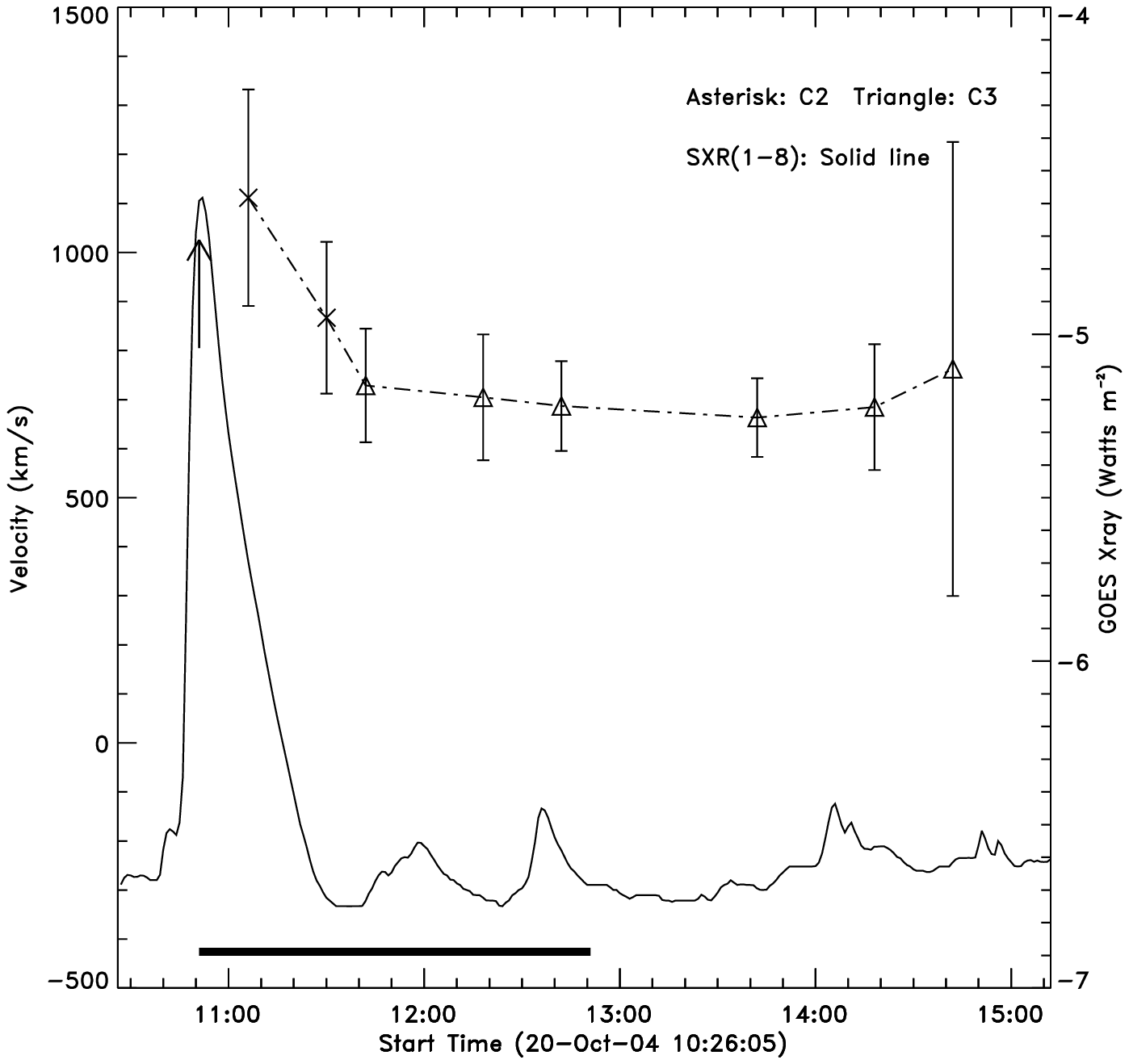} \caption{The velocity
evolution of 2004 October 20 CME (broken line with symbols) and the
temporal profile of the GOES SXR flux of the associated flare (solid
line). The CME has a negative post-impulsive-phase acceleration and
the flare is of short-decaying.\label{fig3}}
\end{figure}

\begin{figure}
\epsscale{1.00} \plotone{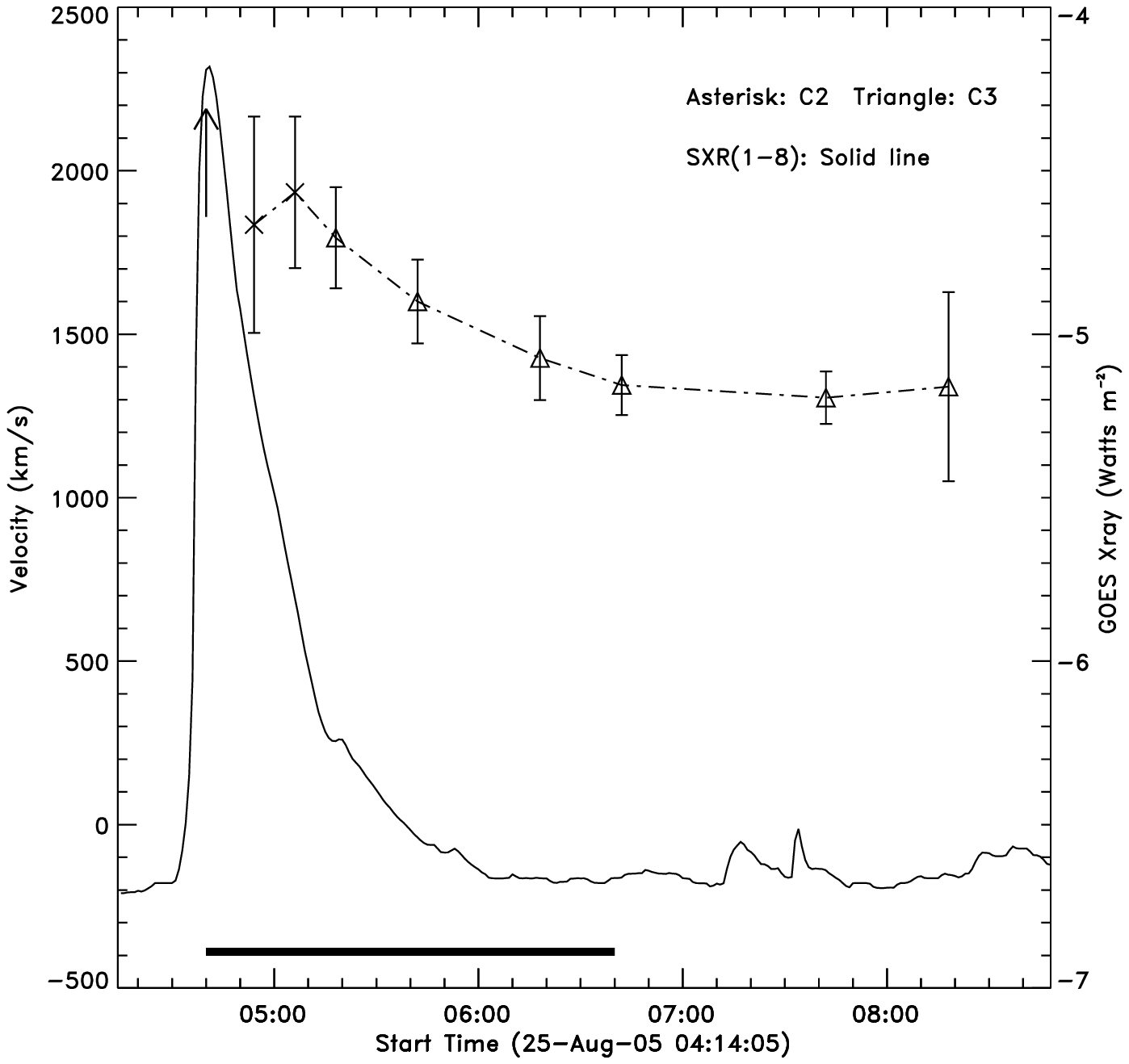} \caption{The velocity
evolution of 2005 August 25 CME (broken line with symbols) and the
temporal profile of the GOES SXR flux of the associated flare (solid
line). The CME has a negative post-impulsive-phase acceleration and
the flare is of short-decaying.\label{fig4}}
\end{figure}

\begin{figure}
\epsscale{1.0} \plotone{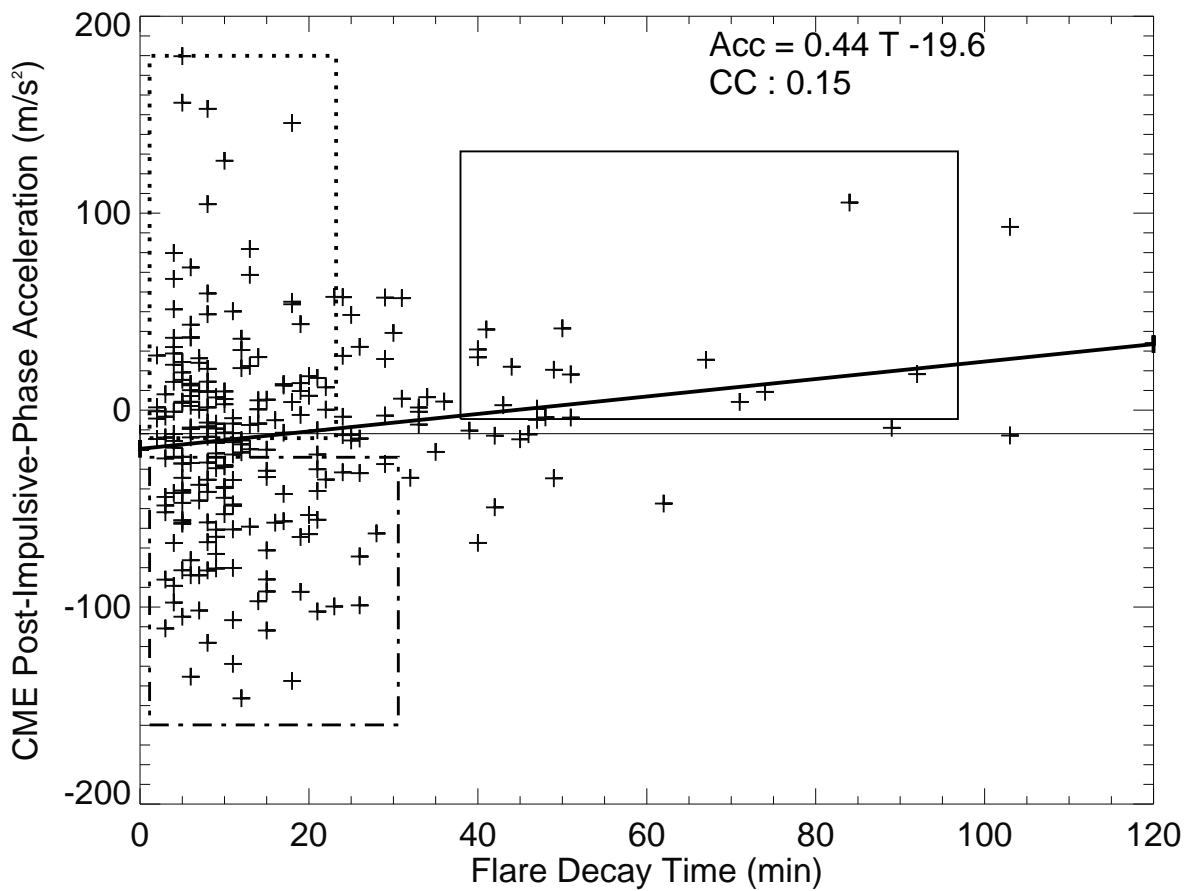} \caption{Scattering plot of
CME post-impulsive-phase acceleration vs. flare decay time for all
the 247 CME-flare events studied in this paper. The solid bold line
denotes the linear fit to the points. The solid thin line indicates
the average post-impulsive-phase acceleration of all events. The
three rectangles group events into different types: acceleration
CMEs with long-decay flares (A-L) (solid), deceleration CMEs with
short-decay flares (D-S) (dash-dotted), and acceleration CMEs with
short-decay flares (A-S). \label{fig5}}
\end{figure}

\begin{figure}
\epsscale{1.0} \plotone{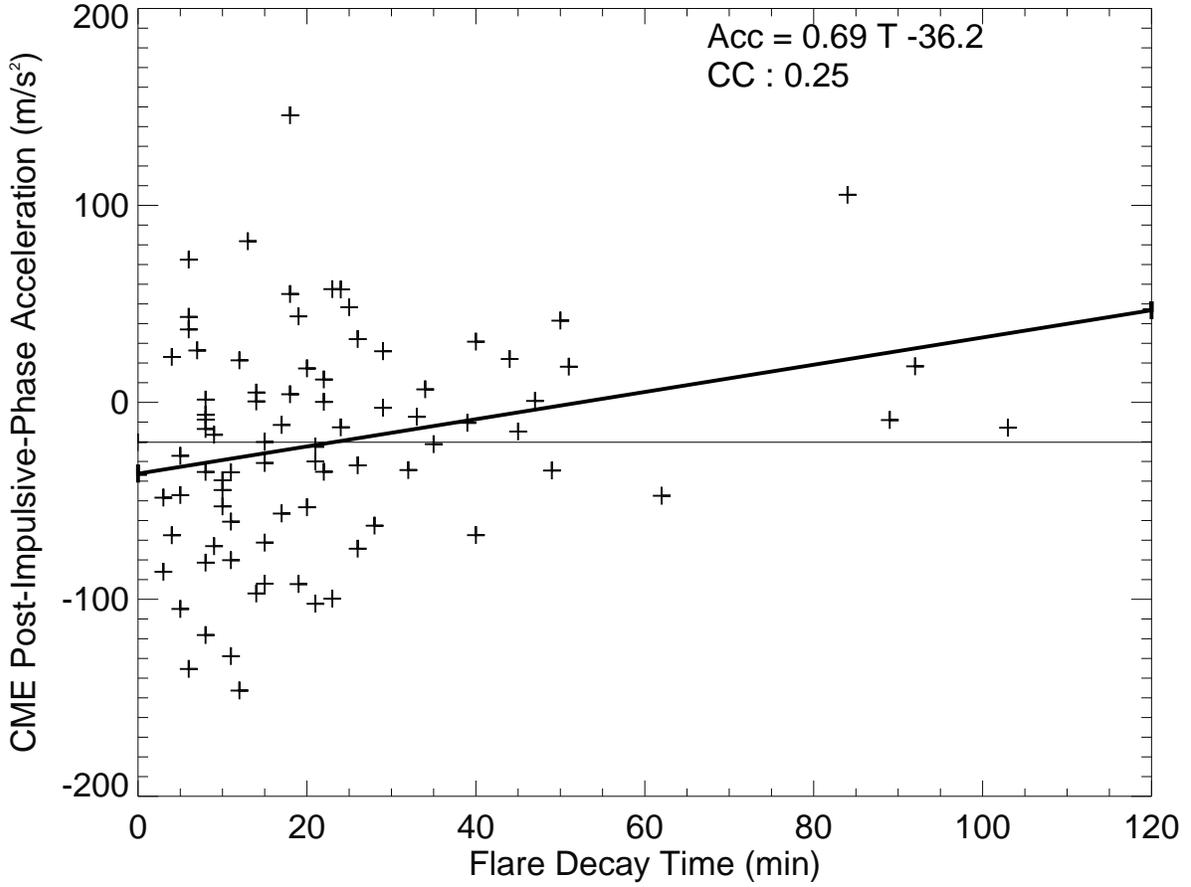} \caption{Scattering plot of
CME post-impulsive-phase acceleration vs. flare decay time for 82
fast and wide CMEs (velocity $\geq$ 800 km s$^{-1}$ and width $\geq$
60$^{\circ}$). The solid bold line denotes the linear fit to the
points. The solid thin line indicates the average
post-impulsive-phase acceleration of these events. \label{fig6}}
\end{figure}

\begin{figure}
\epsscale{1.0}\plotone{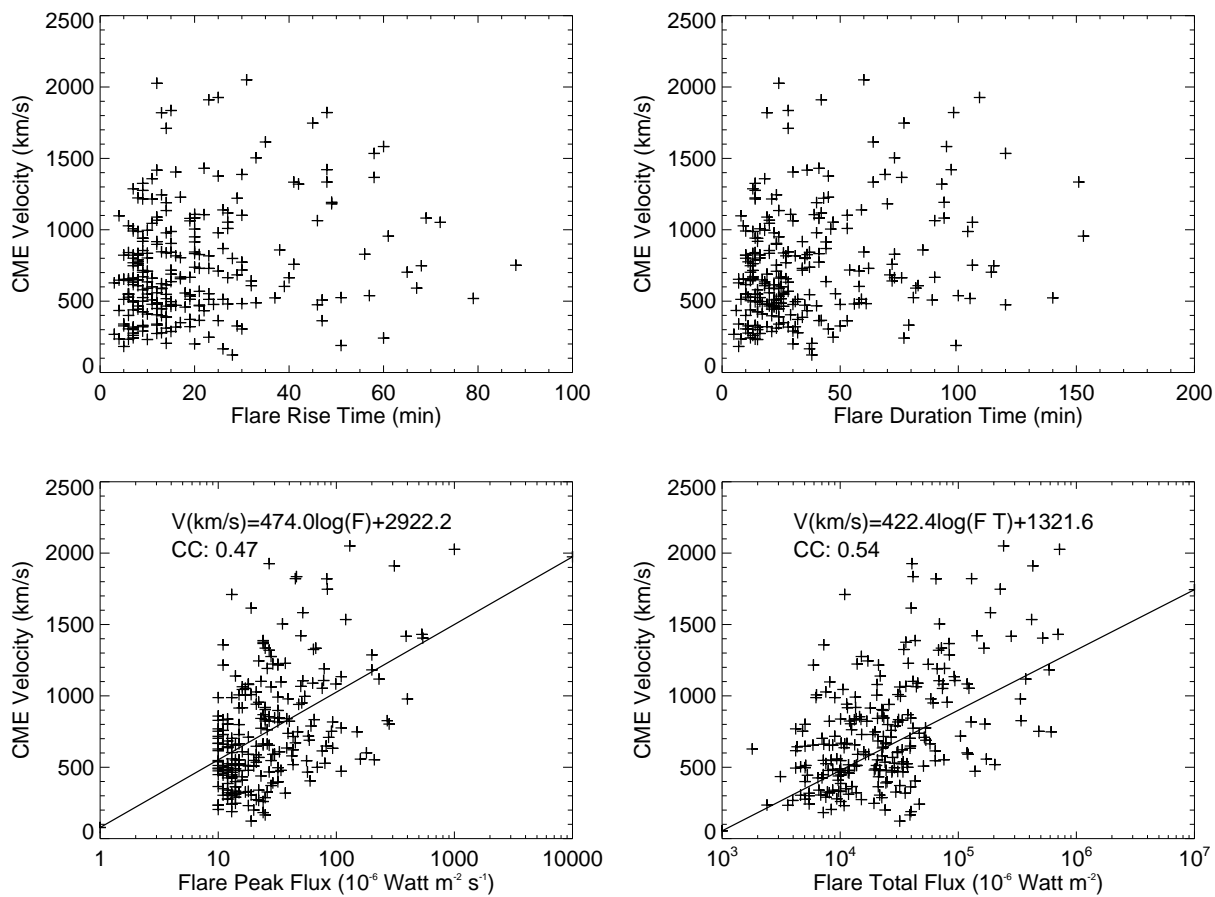} \caption{Scattering plots of CME
velocity versus various flare properties: versus flare rise time
(top-left panel), versus flare total duration (top-right panel),
versus flare peak flux (or magnitude) (bottom-left panel) and versus
flare total flux (or fluence) (bottom-right panel). The linear
correlation fit is carried out in the two bottom panels. Apparently,
the CME velocity-flare total flux has the best correlation with a
coefficient of 0.54. \label{fig7}}
\end{figure}

\begin{figure}
\epsscale{1.0} \plotone{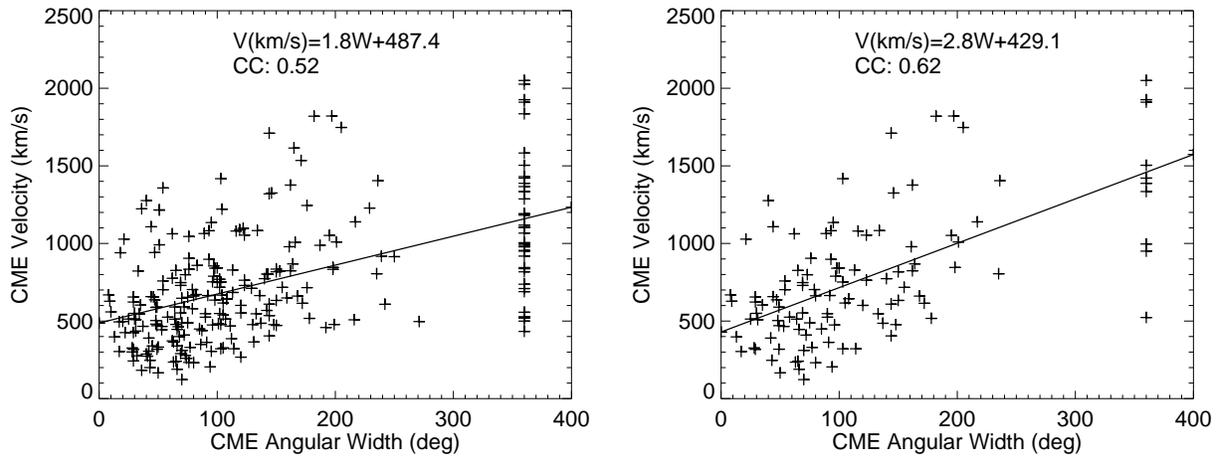} \caption{Scattering plots of CME
velocity versus CME angular width for all the 247 events (left
panel) and 111 limb events (right panel). Solid lines are
linear-correlation fit. CC denotes the correlation
coefficient.\label{fig8}}
\end{figure}

\begin{figure}
\epsscale{0.5} \plotone{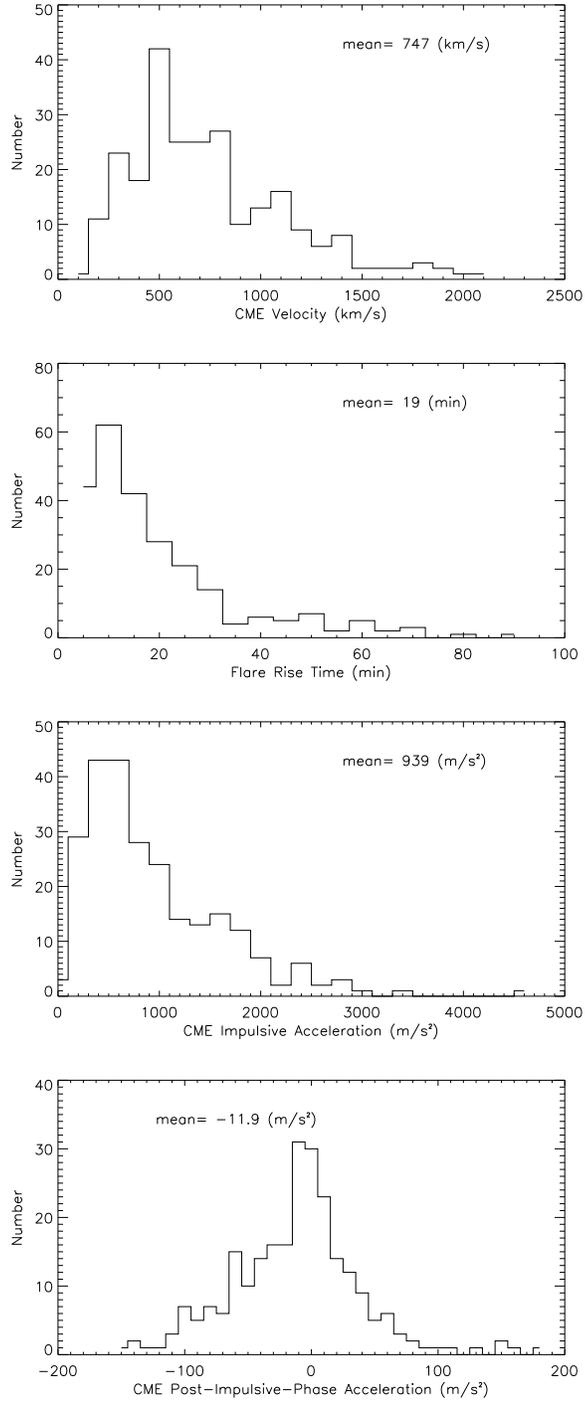} \caption{Histograms of CME and
flare properties.\label{fig9}}
\end{figure}

\clearpage

\begin{table}
\begin{center}
 \caption{Properties of Four Typical CME-flare Events.\label{tb}}
\tabcolsep 2pt
\begin{tabular}{ccccccccc}

\tableline\tableline Event &Property & Velocity  & Im. Acc$\tablenotemark{a}$  &  Post. Acc$\tablenotemark{b} $ (errors) & Rise & Decay & Location & Magnitude \\

 & & (km s$^{-1}$)   &
 (km s$^{-2}$) &  (km s$^{-2}$)&(min) & (min) & (deg) &  \\

\tableline
2001 Sep 24 &A-L &2402  &455&64 ($\pm$58) &66&31 &S16E23 &X2.6\\
2003 Nov 18 &A-L &1821  &632&40 ($\pm$36) &48&50 &S14E89 &M4.5\\

2004 Oct 20 &D-S &763   &1589&--71 ($\pm$37) &8&5 &N11E68 &M2.6\\
2005 Aug 25 &D-S &1327  &2453&--129 ($\pm$41)&9&5&N09E80 &M6.4\\

\tableline
\end{tabular}

\tablenotetext{a}{Magnitude of impulsive acceleration}
\tablenotetext{b}{Magnitude of post-impulsive-phase acceleration}
\end{center}
\end{table}

\begin{table}
\begin{center}
\caption{Statistical Values of CME/flare Properties.\label{tb}}

\begin{tabular}{ccccccc}
\tableline\tableline Parameters &Min &Max & Mean & Med & Mode & Sdev\\
\tableline
Velocity (km s$^{-1}$)             &123  &2050 &747 &657 &427 &391\\
Rise time  (min)                   &3   &88 &19 &14 &12 &15 \\
Im. Acc$\tablenotemark{a}$ (m s$^{-2}$)     &61.8  &4574.1 &939.3 &723.9 &536.8 &693.5 \\
Post. Acc$\tablenotemark{b}$ (m s$^{-2}$) &--146.3  &179.7 &--11.9 &--9.4 &--9.0 &51.0 \\
\tableline
\end{tabular}
\tablenotetext{a}{Magnitude of impulsive acceleration}
\tablenotetext{b}{Magnitude of post-impulsive-phase acceleration}
\end{center}
\end{table}

\end{document}